\documentclass[12pt]{iopart}
\usepackage[utf8]{inputenc}
\usepackage{iopams}
\usepackage{upgreek}
\usepackage{graphicx}
\usepackage{epstopdf}

  \expandafter\let\csname equation*\endcsname\relax
  \expandafter\let\csname endequation*\endcsname\relax
\usepackage{amsmath}

\usepackage{hyperref}

\usepackage{xcolor}
\usepackage[normalem]{ulem}

\begin{document}

\title{Twin-beam real-time position estimation of micro-objects in 3D}

\author{Martin Gurtner and Jiří Zemánek}

\address{Czech Technical University in Prague, Faculty of Electrical Engineering,\\Technická 2, 166 27 Praha 6, Czech Republic}
\ead{martin.gurtner@fel.cvut.cz}
\vspace{10pt}
\begin{indented}
\item[]June 2016
\end{indented}

\begin{abstract}
Various optical methods for measuring positions of micro-objects in 3D have been reported in the literature. Nevertheless, the majority of them are not suitable for real-time operation, which is needed, for example, for feedback position control. In this paper, we present a method for real-time estimation of the position of micro-objects in 3D\footnote{The implementation can be found at \url{http://github.com/aa4cc/twinbeam-measurement}}; the method is based on twin-beam illumination and it requires only a very simple hardware setup whose essential part is a standard image sensor without any lens. The performance of the proposed method is tested during a micro-manipulation task in which the estimated position served as feedback for the controller. The experiments show that the estimate is accurate to within $\mathtt{\sim}3\,\mathrm{\upmu m}$ in the lateral position and $\mathtt{\sim}7\,\mathrm{\upmu m}$ in the axial distance with the refresh rate of 10 Hz. Although the experiments are done using spherical objects, the presented method could be modified to handle non-spherical objects as well.
\end{abstract}



%
\noindent{\it Keywords}: position measurement, twin-beam illumination, digital holography, micro-manipulation, feedback manipulation

%
\submitto{Measurement Science and Technology}

%
%
%

\section{Introduction}

Estimation of positions for micro-objects in 3D is of great interest in many research domains. In microfluidics, a velocity profile of the fluid can be measured by tracking micro-objects suspended in the fluid~\cite{Cheong2009Flow}. For example, blood flow, which could indicate circulatory diseases, can be determined by measuring blood cells' trajectories~\cite{Choi2009ThreeDimensional}. Similarly, an analysis of motion of bubbles in air-water mixture can be carried out~\cite{Tian2010Quantitative}. In microbiology, trajectories of sperm cells can be used to determine their motility~\cite{Caprio20144D}. 

There are many ways how to estimate 3D position of micro-objects. \textit{Confocal microscopy} can be used for 3D position estimation~\cite{Blaaderen1995RealSpace}, but it provides only a very limited time resolution since it involves hardware motion. A variety of methods are based on \textit{digital holography}~\cite{Schnars2015Digital} where the position is estimated computationally; for a review, see~\cite{Yu2014Review}. These methods are either based on fitting the micro-object's hologram to a model describing the appearance of the hologram parametrized by the \textit{axial distance}~\cite{Cheong2009Flow,Viramontes20063D}, or on back-propagation of the hologram~\cite{Lee2007Holographic,Dubois2006Focus}. Fitting the observed holograms to the model provides very accurate estimates of the position (up to nanometer resolution), but it is computationally demanding. In addition, the holograms have to be captured with very high resolution; this is usually achieved by an objective lens which results in reduced observable area. Back-propagation allows us to estimate the axial distance of a micro-object by identification of the distance for which the back-propagated hologram fits the image of the micro-objects. The back-propagation itself is not computationally demanding, but---with the exception of the method described in~\cite{Bouchal2014NonIterative}---it has to be carried out several times. Another approach is to use multiple light sources and subsequently illuminate the micro-objects under different angles~\cite{Su2010MultiAngle,Memmolo2011TwinBeams}. Then the individual micro-object's ``shadows'' on the image sensor are shifted with respect to each other; this shift corresponds to the axial distance of the micro-object.

The majority of methods estimating the position of micro-objects are intended for an analysis of the motion and rely on off-line processing of the recorded data. However, when it comes to feedback position control, one needs to know the positions of the manipulated micro-objects in real-time. Hence, we were motivated to develop a novel method especially suitable for real-time processing and micro-manipulation applications. This method is based on twin-beam illumination and it needs only a very cost-effective and compact hardware setup. The setup consists of two light sources simultaneously illuminating the micro-objects and a standard image sensor (no lens is necessary) capturing the ``shadows'', or more precisely interference patterns, from the micro-objects. The position of a micro-object is computationally estimated from the lateral shift of the corresponding interference patterns. The presented method is tested using a reference measurement from another camera during a micro-manipulation task.

\section{Hardware setup}
\begin{figure*}[!t]
  \centering
  \includegraphics[width=\textwidth]{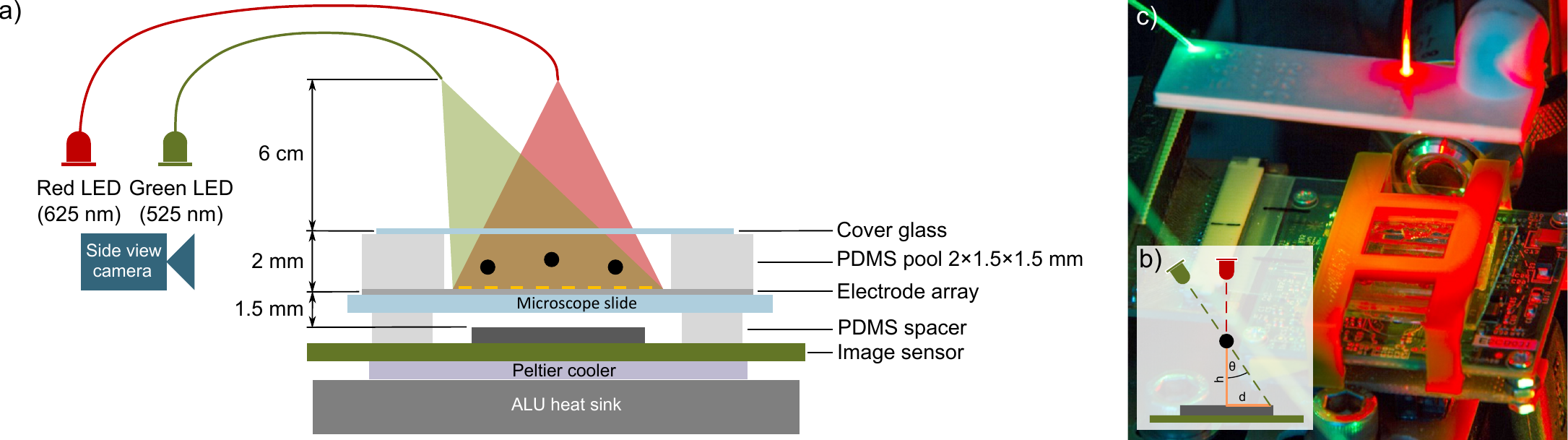}
  \caption{Diagrams (a) of the hardware setup and (b) of the working principle. Figure (c) displays a photo of the hardware setup. PDMS is an abbreviation of polydimethylsiloxane.}
  \label{Fig:hwArrangement}
\end{figure*}

Before we delve into the description of the proposed method, we describe the hardware setup (see figure~\ref{Fig:hwArrangement}). The objects to be tracked are polystyrene spherical micro-objects of diameter $50\,\mathrm{\upmu m}$ that are suspended in water contained in a $2\,\mathrm{mm}$ deep pool above an electrode array. The micro-objects are manipulated through the phenomenon known as \textit{dielectrophoresis}---application of different potentials on the electrodes generates a force acting on the micro-objects~\cite{Morgan2003Ac}. Light sources are red (625 nm) and green (525 nm) LEDs that are butt-coupled to plastic optical fibers ($500\,\mathrm{\upmu m}$ in diameter). The tips of the fibers are placed so, that the light from the red LED illuminates the micro-objects from above and the light from the green LED falls under approximately $45^\circ$. The tips of the fibers are approximately $6\,\mathrm{cm}$ above the pool. Since the light is partially coherent---spatially due to the diameter of the optical fibers and temporally due to the bandwidth of the LEDs---it forms interference patterns on the image sensor (e-Con Systems, See3CAM\_10CUG, 1.3 megapixels, $3.75\,\mathrm{\upmu m}$ pixel size) which is approximately $1.5\,\mathrm{mm}$ below the micro-objects. The image sensor is cooled by a Peltier cooler to avoid heating-up the water in the pool, which would cause undesired heat-driven currents. In our case, the \textit{lateral position} of the micro-objects is restricted by the size of the electrode array to approximately $1.5\times1.5\,\mathrm{mm}$ and the axial distance by dielectrophoresis to maximum levitation height of $200\,\mathrm{\upmu m}$ above the electrode array. For the calibration of the method and validation purposes, there is also a side-view camera that allows us---in a very limited depth of field---to see the micro-objects in the pool from aside.

\section{Working principle}
\begin{figure*}[t]
	\centering
	\includegraphics[width=0.7\textwidth]{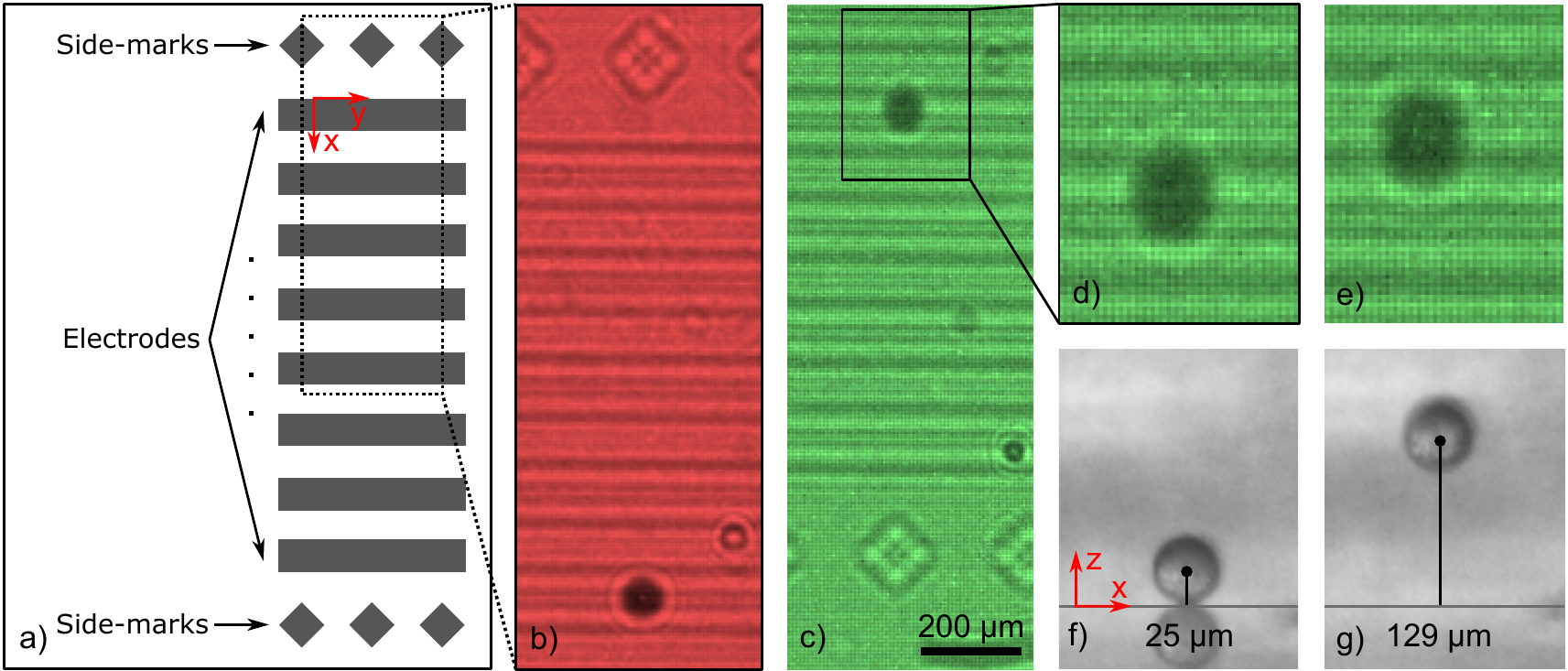}
	\caption{Demonstration of the principle of the proposed method. (a) Illustration of the electrode array with the side-marks. Images (b) and (c) are red and green channels of a cut-out of a captured image and they contain only interference patterns from perpendicular and oblique illumination, respectively. The images contain shifted interference patterns from the electrode array and one micro-object. (d-e) Blown-up regions of the green channel for a micro-object located at different levitation heights as it is shown from the side view in (f-g). The interference pattern of the micro-object (d) is shifted with respect to (e) and this shift corresponds to the difference in the axial distance of the micro-object.}
	\label{fig:colorSepar_heightPrinciple}
\end{figure*}

The principle of the proposed method is depicted in figure~\ref{Fig:hwArrangement}. As the micro-objects are illuminated by one source from the top and by the other one from the side, there are two interference patterns on the image sensor under each micro-object. These two interference patterns are laterally shifted with respect to each other and this shift corresponds to the axial distance of the micro-object. Because the wavelengths of the LEDs were chosen so that they match the peaks in the sensitivity of the red and green channels of the image sensor, the red and green channels contains only interference patterns from the perpendicular and oblique illumination, respectively (see figure~\ref{fig:colorSepar_heightPrinciple}). If we assume that the micro-objects are illuminated by planar waves and neglect the refraction of light, the dependence of the axial distance of the micro-objects on the lateral shift is simply given by
\begin{equation}
\label{eq:height_lateraShift}
	h = d \frac{1}{\tan\theta},
\end{equation}
where $h$ is the axial distance of a micro-object from the image sensor, $d$ is the lateral shift of its interference patterns in the captured image and $\theta$ is the angle of the oblique incidence (see figure~\ref{Fig:hwArrangement}(c)).

Nevertheless, the tips of the optical fibers behave more like sources of spherical waves, and the refraction of light clearly occurs because the light propagates through several different media on the way from the tips to the image sensor. The assumption of planar wave illumination is a good approximation if the tips of the optical fibers are sufficiently far away from the micro-objects and the lateral shift of the interference patterns is measured close enough to the micro-objects---ideally in the same medium to avoid the additional refraction of light. However, putting farther the light sources would enlarge the hardware setup and would require more energy for the same intensity of light incident on the image sensor. Putting the image sensor closer to the micro-objects is also rather difficult because that would mean making the electrode array and the supporting microscope glass thinner. To overcome this, side-marks are placed along the electrodes (see figure~\ref{fig:colorSepar_heightPrinciple}) which allows us to find a transformation from the image coordinate system to the electrode array coordinate system. This way we can effectively measure the lateral shift at the level of the electrode array, which is very close to the micro-objects. Hence, we significantly reduced the influence of refraction of light and the influence of the non-planar illumination.

We assume that the transformations from the red and green channels (image coordinate systems) to the electrode array coordinate system can be described by a \textit{projective transformation}~\cite{Hartley2004Multiple}. That is, for the red channel, image coordinates $(x_\mathrm{im}, y_\mathrm{im})$ are transformed to electrode array coordinates $(x_\mathrm{el}, y_\mathrm{el})$ by the following relation
\begin{equation}
  \begin{bmatrix}
    x_\mathrm{el}w \\
    y_\mathrm{el}w \\
    w
  \end{bmatrix}
  =
  \mathsf{H}_\mathrm{R}
  \begin{bmatrix}
    x_\mathrm{im} \\
    y_\mathrm{im} \\
    1 
  \end{bmatrix},
\end{equation}
where $\mathsf{H}_\mathrm{R}\in\mathbb{R}^{3 \times 3}$. The same relation applies for the green channel, only the transformation matrix differs.

This is where the side-marks are useful; in order to determine the parameters of the projective transformation, one needs at least four pairs of corresponding points in both coordinate systems~\cite{Hartley2004Multiple}. Relative positions of side-marks in the electrode array coordinate system are known, and positions of several side-marks in the image coordinate systems (for robustness, more than the needed four) are provided by a user. Therefore, such transformation parameters can be found and we can transform the positions of interference patterns to the electrode array coordinate system and measure their lateral shift there.

Now, we identify the precise locations of individual interference patterns and pair the patterns corresponding to the same micro-particle in the red and green channels. At the initial stage, approximate positions of the interference patterns from the perpendicular illumination (red channel) are provided by the user. Since the axial distance of the micro-objects is limited to a very narrow range, the mutual position of the interference patterns from the perpendicular and oblique illumination differs only slightly (up to 14 pixels). Thus, given the position of interference patterns from the perpendicular illumination the approximate position of the corresponding interference patterns from the oblique illumination can be calculated. To refine the approximate positions, the color channels are back-propagated to a distance where the interference patterns focus to a point. We do this because it is easier to determine a precise location of a focused point than of a larger interference pattern. The back-propagation is carried out by calculating the Rayleigh-Sommerfeld diffraction integral~\cite{Goodman1996Introduction} which is numerically done by the following relation
\begin{equation}
        I_z(x_\mathrm{im}, y_\mathrm{im}) = \mathcal{F}^{-1}\left\{ H_{-z}(f_x, f_y) \mathcal{F}\left\{ I(x_\mathrm{im}, y_\mathrm{im}) \right\}\right\},
\end{equation}
where $(x_\mathrm{im}, y_\mathrm{im})$ are the image coordinates, $(f_x, f_y)$ are the spatial frequencies, $I$ is the original image, $I_z$ is the image back-propagated to a distance $z$, $\mathcal{F}$ and $\mathcal{F}^{-1}$ are Fourier and inverse Fourier transformations, respectively, and
\begin{equation}
        H_z(f_x, f_y) =
        	\begin{cases}
                   \exp\left( i 2 \pi z \frac{n}{\lambda} \sqrt{1 - \left(\frac{\lambda f_x}{n}\right)^2 - \left(\frac{\lambda f_y}{n}\right)^2}\right), & \sqrt{f_x^2+f_y^2}\! \leq\!\! \frac{n}{\lambda}, \\
                0,       & \text{otherwise},
            \end{cases}
\end{equation}
is Fourier transform of the Rayleigh-Sommerfeld propagator, where $\lambda$ is the wave length of the illuminating light and $n$ is the refractive index. Despite the fact that the light propagates through several media on the way from the microparticle to the image sensor, we use a fixed value of the refractive index to make the back-propagation computationally faster.

\begin{figure*}[!t]
	\centering
	\includegraphics[width=0.75\textwidth]{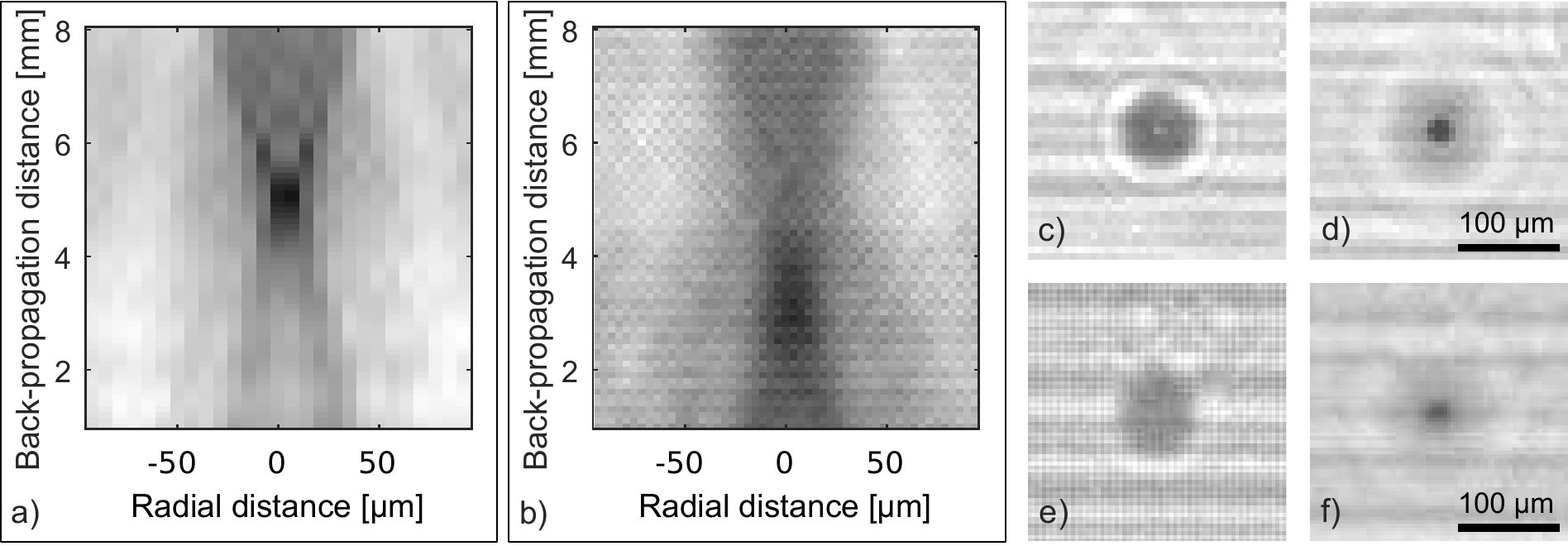}
	\caption{Back-propagation of an interference pattern from a micro-sphere: Images (a) and (b) show the dependence of radial intensity of the interference pattern from perpendicular and oblique illumination, respectively, on the back-propagation distance. Images (c) and (e) display raw interference patterns from perpendicular and oblique illumination, respectively, and (d) and (f) shows their back-propagation to a distance where they are focused to a point. Units of back-propagation distance correspond to the fixed refractive index.}
	\label{fig:intPatProp}
\end{figure*}

The back-propagation is illustrated in figure~\ref{fig:intPatProp}. For each light source, we can separately fix a back-propagation distance for which all the interference patterns are focused to a point, no matter where the micro-objects are located.

The position of focused interference patterns is estimated by computing the center of mass of a small region around the approximate position (the position in the previous frame or, at the initial stage, the position given by the user). To make the estimate more accurate, the center of mass is calculated for regions of successively smaller sizes to eliminates the influence of the surrounding specks.

What remains to be done is to identify the constant $1/{\tan\theta}$ in~\eqref{eq:height_lateraShift}. For this purpose, we use the side-view camera in the hardware setup. The side-view camera allows us to measure the levitation height (axial distance) of micro-objects within a limited depth of field directly (see~figure~\ref{fig:colorSepar_heightPrinciple} (f-g)). We manipulate a micro-object to several levitation heights and measure those heights and lateral shifts of the corresponding interference patterns. The constant $1/{\tan\theta}$ is then identified by fitting~\eqref{eq:height_lateraShift} to the set of measured points. This calibration has to be done only once for the hardware setup and then the side-view camera is no longer needed.

\section{Experimental results}
\begin{figure*}[t]
  \centering
  \includegraphics[width=\textwidth]{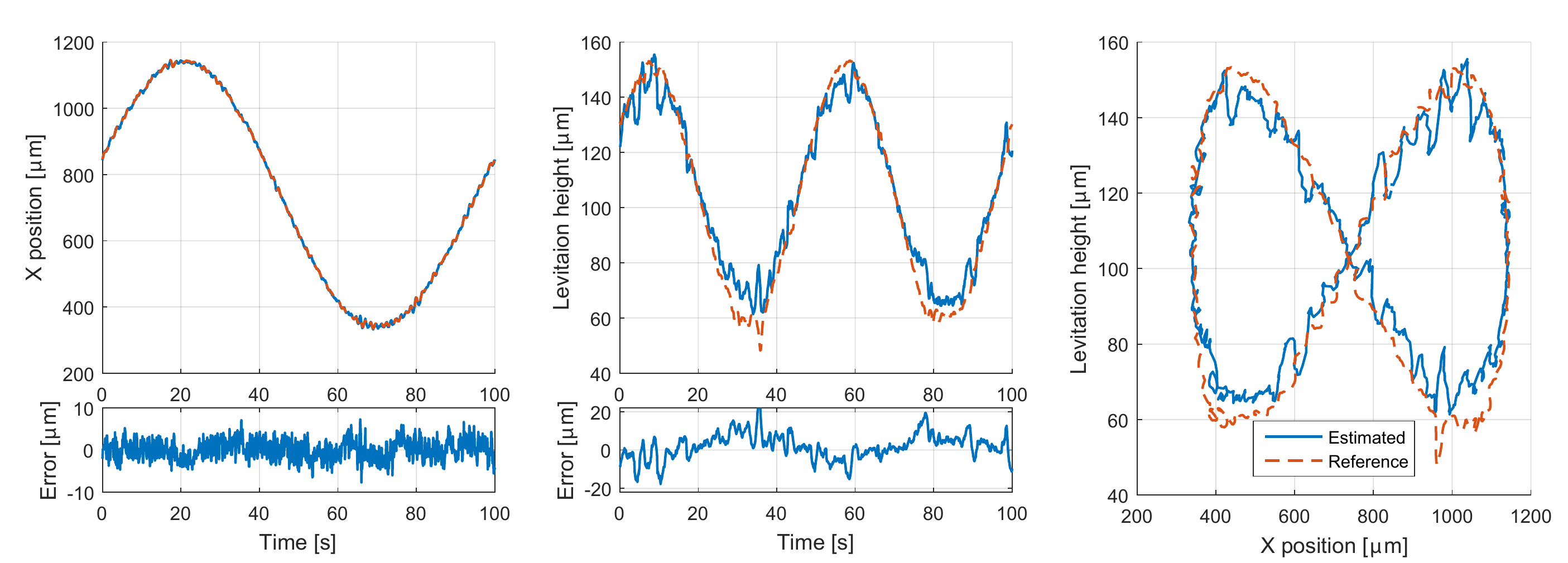}
  \caption{Comparison of the estimated positions estimated by the proposed method with the reference measurements obtained by the side-view camera.}
  \label{fig:accuracyAssesment}
\end{figure*}

To validate the performance of the proposed method we manipulate a micro-object (polystyrene microsphere with $50\,\mathrm{\upmu m}$ in diameter) along a figure-eight trajectory and compare the position estimated by the proposed method with the reference measurement obtained from the side-view camera with accuracy $\mathtt{\sim} 0.25\,\mathrm{\upmu m}$. The proposed method is implemented in Simulink. The estimation is carried out in real-time at $10\,\mathrm{Hz}$ on an ordinary PC (Intel Core i7, 8 GB RAM) and it is used in the feedback loop of the control algorithm described in~\cite{Zemanek2015Feedbacka}. The estimation algorithm itself takes only $40\,\mathrm{ms}$ and the remaining $60\,\mathrm{ms}$ is taken by the control algorithm and the execution overhead. The comparison is displayed in the form of graphs in figure~\ref{fig:accuracyAssesment} and in the form of video (containing also the side-view) available at \url{https://youtu.be/150__OV3aUk}. The side-view camera enables us to measure only one coordinate of the lateral position, but from the method of estimation, the estimate in the other coordinate has necessarily the same accuracy. The standard deviation of the error in $x$-coordinate is $2.41\,\mathrm{\upmu m}$ ($0.6\,\mathrm{px}$) and in the levitation height $6.64\,\mathrm{\upmu m}$. Even though the experiment is performed with only one micro-object, there is no obstacle preventing exploiting the proposed method for tracking of several micro-objects. Such an experiment is shown in a video available at \url{https://youtu.be/vbNSIDCg4Bg}.

\section{Discussion and conclusion}
We have developed a simple, novel method for real-time estimation of the position of spherical micro-objects. The method requires only a very simple, cost-effective and compact hardware setup. We demonstrated the accuracy to be within $\mathtt{\sim} 3\,\mathrm{\upmu m}$ in the lateral position and $\mathtt{\sim} 7\,\mathrm{\upmu m}$ in the axial distance. Since the accuracy depends on precise localization of the interference patterns, it can be improved by using an image sensor with smaller pixels, but this usually reduces the observable area. Furthermore, we successfully used the method for real-time manipulation of a micro-object. Despite the fact that the method is developed for transparent spherical micro-objects, it can be potentially extended to track non-spherical and/or opaque micro-objects as well. The only thing that would have to change is the localization procedure for the interference patterns because they might not focus to a point anymore. Concerning limitations of the proposed method, if the micro-objects are in contact or located at the same lateral position (they lie along the same axial line) it might be difficult to track them with the current system.

\ack{Acknowledgments}
This research was funded by the Czech Science Foundation within the project P206/12/G014 (Centre for advanced bioanalytical technology, http://www.biocentex.cz).

\section{References} 
\label{sec:references}
\bibliographystyle{unsrt}
\bibliography{References}

\end{document}